\documentclass[10pt, conference, compsocconf]{IEEEtran}
\usepackage{times}

\usepackage{graphicx}
\graphicspath{{figures/}}
\usepackage{url}
\usepackage{subfigure}
\usepackage{tikz}
\usepackage{xspace}
\usepackage[small,it]{caption} 
\usepackage[ruled,vlined]{algorithm2e}
\usepackage{fourier} 

\newcommand{\ignore}[1]{{}}
\newtheorem{lemma}{Lemma}

\newtheorem{corollary}{Corollary}

\newcommand{\qed}{\nobreak \ifvmode \relax \else
  \ifdim\lastskip<1.5em \hskip-\lastskip
  \hskip1.5em plus0em minus0.5em \fi \nobreak
  \vrule height0.75em width0.5em depth0.25em\fi
}
\newcommand{\eg}{\textit{e.g.}\xspace}
\newcommand{\ie}{\textit{i.e.}\xspace}
\newcommand{\happensbefore}{$\rightarrow$}

\newcommand{\Section}[1]{
  \section{\protect{#1}}
}
\newcommand{\SubSection}[1]{
  \subsection{\protect{#1}}
}

\newcommand\chronogramme[1]{
      \foreach \x in { 0,...,#1 } {
        \draw[arrows=->,thick] ( \x*4, 0 ) -- ( \x*4, -11 );
        \node (P\x) at ( \x*4, 1 ) {P\x};
	\fill (\x*4, 0) circle(4pt);
	\noeud{(\x*4,0)}{PP\x}
      }
      \tikzstyle{frame} = [draw,thick,-latex]
}

\newcommand\noeud[2]{
	\path #1 coordinate (#2);
	\fill (#2) circle(4pt);
}

\newsavebox{\maboite}
\newlength{\Lmaboite}

\newcommand\noeudMatrice[3]{
	\sbox{\maboite}{#3}
	\settowidth{\Lmaboite}{\usebox{\maboite}}
	\node at (#1)[#2] {\parbox{0.5cm}{\setlength{\baselineskip}{0pt}#3}};
}

\title{A Model for Coherent Distributed Memory\\
For Race Condition Detection} 

\author{\IEEEauthorblockN{
\begin{tabular}{cc}
Franck Butelle & Camille Coti  \\
Franck.Butelle@lipn.univ-paris13.fr &
Camille.Coti@lipn.univ-paris13.fr \\
\end{tabular}
}
\IEEEauthorblockA{LIPN, CNRS-UMR7030, Universit\'{e} Paris 13, F-93430 Villetaneuse,
  France
}
}

\date{}

\begin{document}

\maketitle

\begin{abstract}
We present a new model for distributed shared memory systems, based
on remote data accesses.
Such features are offered by network interface cards
that allow one-sided operations, remote direct memory access and
OS bypass.
This model leads to new interpretations of distributed algorithms
allowing us to propose an innovative detection
technique of race conditions only based on logical clocks.
Indeed, the presence of (data) races in a parallel program 
makes it hard to reason about and is usually considered as a bug. 
\end{abstract}

\Section{Introduction}


The \emph{shared-memory model} is a convenient model for programming
multiprocessor applications: all the processes of a parallel
application running on different processors have access to a common
area of memory. Another possible communication model for distributed systems is
the \emph{message-passing model}, in which each process can only access its own
local memory and can send and receive message to other processes.

The message-passing model on distributed memory requires to move data
between processes to make it available to other
processes. Under the shared-memory model, 
all the processes can read or write at any address of the shared memory. 
The data is \emph{shared} between all the processes.

One major drawback of the shared-memory model for practical situations
is its lack of scalability. A direct implementation of shared memory
consists in plugging several processors / cores on a single
motherboard, and letting a single instance of the operating system
orchestrate the memory accesses. Recent blades for supercomputers
gather up to 32 cores per node, Network on Chip (NoC) systems embed 80
cores on a single chip: although the ``many-core'' trend increased drastically
the number of cores sharing access to a common memory bank,
it is several orders of magnitude behind current supercomputers: in
the Top 500\footnote{\url{http://www.top500.org}} list issued in
November 2010, 90\% of the systems have 1K to 16K cores each.

The solution to benefit from the flexibility and convenience of shared
memory on distributed hardware is \emph{distributed
  shared memory}. All the processes have access to a \emph{global
  address space}, which is distributed over the processes. The memory
of each process is made of two parts: its \emph{private} memory and
its \emph{public} memory. The private memory area can be accessed from
this process only. The public memory area can be accessed remotely
from any other process without notice to the process that maps this
memory area physically.

The notion of global address space is a key concept of parallel
programming languages, such as UPC~\cite{UPC},
Titanium~\cite{YSPMLKHGGCA98} or Co-Array
Fortran~\cite{NR98}. The programmer sees the global memory space as if
it was actually shared memory. The compiler translates accesses to
shared memory areas into remote memory accesses. The run-time
environment performs the data movements. As a consequence, programming
parallel applications is much easier using a parallel language than
using explicit communications (such as MPI~\cite{Forum94}): data
movements are determined by the compiler and handled automatically by
the run-time environment, not by the programmer himself.

The memory consistency model followed by these languages, such as the
one defined for UPC~\cite{KBW04}, does not define a global order of
execution of the operations on the public memory area. As a
consequence, a parallel program defines a set of possible executions
of the system. The events in the system may happen in different orders
between two consecutive executions, and the result of the computation
may be different. For example, if a process writes in an area of
shared memory and another process reads from this location. If the
writer and the reader are two different processes, the memory
consistency model does not specify any kind of control on the order in
which these two operations are performed. Regarding whether the reader
reads before or after the data is written, the result of the writing
may be different.

In this paper, we introduce a model for distributed shared memory that
represents the data movements and accesses between processes at a
\emph{low} level of abstraction. In this model, we present a mechanism
for detecting race conditions in distributed shared memory systems. 

This model is motivated by Remote Direct Memory Access capabilities of
high-speed, low-latency networks used for high-performance computing,
such as the InfiniBand standard\footnote{\url{http://www.infinibandta.org/}} or
Myrinet\footnote{\url{http://www.myri.com}}. 

The remainder of this paper is organized as follows. In
section~\ref{sec:related}, we present an overview of previous models
for distributed shared memory and how consistency and coherency has
been handled in these models. In section~\ref{sec:model} we present
our model for distributed shared memory and how it can be related to
actual systems. In section~\ref{sec:algo} we present how race
conditions can be represented in this model, and we propose an
algorithm for detecting them.


\Section{Previous work}
\label{sec:related}

Distributed shared memory is often modeled as a large cached
memory~\cite{HHSRH95}. The local memory of each node is considered as a
cache. If a process running on this node tries to access some data, it
gets it directly if the data is located in its cache. Otherwise, a
page fault is raised and the distributed memory controller is called
to resolve the localisation of the data. Once the data has been
located (\ie, once the local process knows on which process it is
physically located and at which address in its memory), the
communication library performs a point-to-point communication to
actually transfer the data.

In \cite{Lamport79}, L. Lamport defines the notion of
\emph{sequential consistency}: on each process, memory requests are
issued in the order specified by the program. However, as stated by
the author, sequential consistency is not sufficient to guarantee
correct execution of multiprocessor shared memory programs. The
requirement to ensure correct ordering of the memory operations in
such a distributed system is that a single FIFO queue treats and
schedules memory accesses from all the processes of the system. 

Maintaining the coherence of cache-based distributed shared memory can then be
considered as a cache-coherency problem. \cite{LH86} describes several
distributed and centralized memory managers, as well as how coherence can be
maintained using these memory managers.

However, in a fully distributed system (\ie, with no central memory manager) 
with RDMA and OS bypass capabilities, a process can actually access another
process's memory without help from any memory manager. In parallel languages 
such as UPC~\cite{UPC}, Titanium~\cite{YSPMLKHGGCA98} and Co-Array
Fortran~\cite{NR98}, data locality (\ie, which process holds the data in its
local memory) is resolved at compile-time. 

The MPI-2 standard~\cite{MPI-2} defines remote memory access operations. The
MARMOT error checking tool~\cite{KR06} checks correct usage of the
synchronization features provided by MPI, such as fences and windows.



\Section{Memory and communication model}
\label{sec:model}

In this section, we define a model for \emph{distributed shared memory}. This
model works at a lower level than most models described previously in the
literature. It considers inter-process communications for remote data accesses. 

\SubSection{Distributed shared memory model}
\label{sec:model:dsm}

In many shared-memory models that have been described in the 
literature~\cite{attiya98, Dolev00, Tel94}, pairs of processors communicate 
using registers where they read and write data. Distributed shared
memory cannot use registers between processors because they are
\emph{physically} distant from each other; like message-passing systems, they
can communicate only by using an interconnection network.

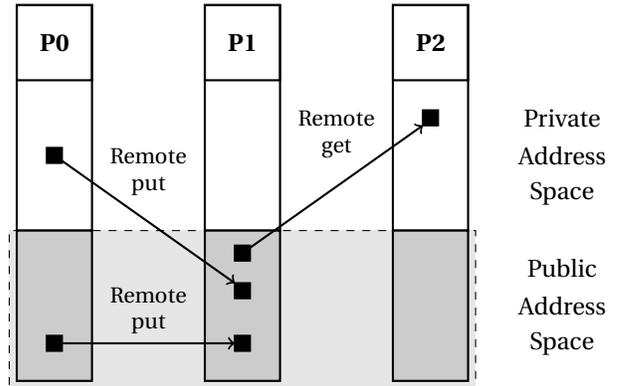
\begin{figure}[ht]
  \centering
  \begin{tikzpicture}[scale=.5]
    \path[draw, dashed, fill=gray!20] ( -.2, -6 ) rectangle ( 12.2, -10.2 );
    
    \foreach \x in { 0, 1, 2 } {
      \path[draw,thick] ( \x*5, 0 ) rectangle ( \x*5+2, -6 );
      \path[draw,thick] ( \x*5, 0 ) rectangle ( \x*5+2, -2 );
      \node at ( \x*5+1, -1 ) {\textbf{P\x}};
      \path[draw,thick,fill=gray!40] ( \x*5, -6 ) rectangle ( \x*5+2, -10 );
    }
    
    \node at ( 14.5, -3 ) {Private};
    \node at ( 14.5, -4 ) {Address};
    \node at ( 14.5, -5 ) {Space};
    
    \node at ( 14.5, -7 ) {Public};
    \node at ( 14.5, -8 ) {Address};
    \node at ( 14.5, -9 ) {Space};
    
    \path[draw,thick,fill=black] ( 10.8, -3.2 ) rectangle ( 11.2, -2.8 );
    \path[draw,thick,fill=black] ( 6.2, -6.4 ) rectangle ( 5.8, -6.8 );
    \draw[ ->, thick] ( 6, -6.6 ) -- ( 10.8, -3.2 );
    \node at ( 8.5, -3 ) {\small Remote};
    \node at ( 8.5, -3.8 ) {\small get};
    
    \path[draw,thick,fill=black] ( .8, -4.2 ) rectangle ( 1.2, -3.8 );
    \path[draw,thick,fill=black] ( 6.2, -7.4 ) rectangle ( 5.8, -7.8 );
    \draw[ ->, thick] ( 1, -4 ) -- ( 5.8, -7.4 );
    \node at ( 3.5, -4 ) {\small Remote};
    \node at ( 3.5, -4.8 ) {\small put};

    \path[draw,thick,fill=black] ( .8, -9.2 ) rectangle ( 1.2, -8.8 );
    \path[draw,thick,fill=black] ( 6.2, -9.2 ) rectangle ( 5.8, -8.8 );
    \draw[ ->, thick] ( 1, -9 ) -- ( 5.8, -9 );
    \node at ( 3.5, -7.7 ) {\small Remote};
    \node at ( 3.5, -8.5 ) {\small put};
  \end{tikzpicture}
  \caption{\label{fig:memorymodel}Memory organization of a three-processor
    distributed shared memory system.}
\end{figure}

Figure~\ref{fig:memorymodel} depicts our model of organization of the public and
private memory in a multiprocessor system. In this model, each processor maps
two distinct areas of memory: a \emph{private} memory and a \emph{public}
memory. The private memory can be accessed from this processor only. 

The public address space is made of the set of all the public memories
of the processors (the \emph{Global Address Space}). Processors can copy data
from/to their private memory and the public address space, regardless of data
locality. 

Public memory can be accessed by any processor of the application, in
\emph{concurrent} read and write mode. In particular, no distinction is made
between accesses to public memory from a remote process and from the process
that actually maps this address space.

The compiler is in charge with data locality, \ie, putting shared data in the
public memory of processors. For instance, if a data $x$ is defined as
shared by the programmer, the compiler will decide to put it into the memory of
a processor $P$. Instead of accessing it using its address in the local memory,
processors use the processor's name and its address in the memory of this
processor. This couple $(processor\_name, local\_address)$ is the addressing
system used in the global address space. The compiler also makes the address
resolution when the programmer asks a processor to access this shared data $x$.

In addition, since NICs (Network Interface Controllers) 
are in charge with memory management in the public
memory space, they can provide \emph{locks} on memory areas. These locks
guarantee exclusive access on a memory area: when a lock is taken by a process,
other processes must wait for the release of this lock before they can access
the data.

\SubSection{Communications}
\label{sec:model:comm}

Processor access areas of public memory mapped by other processors using
point-to-point communications. They use \emph{one-sided communications}: the
process that initiates the communication can access remote data without any
notification on the other processor's side. Hence, a processor $A$ is not aware
of the fact that another processor $B$ has accessed (\ie, read or written) in
its memory. 

Accessing data in another processor's memory is called \emph{Remote Direct
  Memory Access} (RDMA). It can be performed with no implication from the remote 
processor's operating system by specific network interface cards, such as 
InfiniBand and Myrinet technologies. It must be noted that the operating system
is not aware of the modifications in its local shared memory. The SHMEM~\cite{B04}
library, developed by Cray, also implements one-sided operations on top of
\emph{shared} memory. As a consequence, the model and algorithms presented in
this paper can easily be extended to shared memory systems.

\begin{figure}[htb]
  \centering
  \begin{tikzpicture}[scale=.5]
    \foreach \x in { 0, 1, 2 } {
      \node at ( \x*4, 1 ) {P\x};
      \draw [->] ( \x*4, 0 ) -- ( \x*4, -6 );
    }
    \draw[thick] ( 0, -1 ) -- ( 4, -3 );
    \draw[ -latex, thick] ( 4, -3 ) -- ( 0, -5 );
    \node at ( 1, -3 ) {get};
    \draw[ -latex, thick] ( 8, -3 ) -- ( 4, -5 );
    \node at ( 6.5, -2.7 ) {put};
  \end{tikzpicture}
  \caption{\label{fig:model:comm}Remote R/W memory accesses.}
\end{figure}
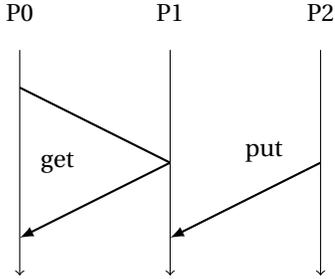

RDMA provides two communication primitives: \emph{put} and \emph{get}. These two
operations are represented in figure~\ref{fig:model:comm}. They are both
\emph{atomic}.

\emph{Put} consists in writing some data into the public memory of another
processor. It involves one message, from the source processor to the destination
processor, containing the data to be written. In figure~\ref{fig:model:comm},
$P2$ writes some data into $P1$'s memory.

\emph{Get} consists in reading some data from another processor's public
memory. It involves two messages: one to request the data, from the requesting
processor to the processor that holds the data, and one to actually transfer the
data, from the processor that holds the data to the requesting processor. In
figure~\ref{fig:model:comm}, $P2$ reads some data from $P1$'s memory.

Communications can also be done within the public space, when data is copied
from a place that has affinity to a process to a place that has affinity to
another process.

\begin{figure}[ht]
  \centering
  \begin{tikzpicture}[scale=.4]
    \footnotesize
    \foreach \x in { 0, 1, 2 } {
      \draw [->] ( \x*5, 0 ) -- ( \x*5, -7.5 );
      \node at ( \x*5, 1 ) {P\x};
    }
    
    \node at ( 7.5, -1 ) {get};
    \draw[ ->, thick] ( 5, -1 ) -- ( 10, -3 );
    \draw[ ->, thick] ( 10, -3) -- ( 5, -5 );
    
    \node at ( 2.5, -1 ) {put};
    \draw[ thick] ( 0, -1 ) -- ( 4.3, -2.5 );
    \draw[ thick] ( 4.3, -2.5 ) -- ( 4.3, -5.6 );
    \draw[ ->, thick] ( 4.3, -5.6 ) -- ( 5, -5.6 );
  \end{tikzpicture}
  \caption{A put operation is delayed until the end of the get operation on the
    same data.}
  \label{fig:model:delayed}
\end{figure}
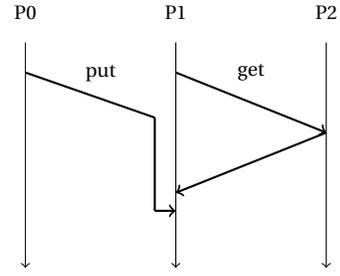

The \emph{get} operation is \emph{atomic} (and therefore, blocking). If a thread gets
some data and writes it in a given place of its public memory, no other thread
can write at this place before the \emph{get} is finished. The second operation is
delayed until the end of the first one (figure~\ref{fig:model:delayed}).

\subsection{Race conditions}
\label{sec:model:rc}

One major issue created by one-sided communications is that several processors
can access a given area of memory without any synchronization nor mutual
knowledge. For example, two processors $A$ and $B$ can write at the same address
in the shared memory of a third processor $C$. Neither $B$ nor $C$ know that $A$
has written or is about to write there.

Concurrent memory accesses can lead to \emph{race conditions} if
they are performed in a totally anarchic way (although some authors
precise \emph{data} race conditions, we will use only "race conditions"
throughout this paper). A race condition is observed when the result
of a computation differs between executions of this computation.
Race condition makes, at least, hard to reason about a program
and therefore is usually considered as a bug.

In the kind of systems we are considering here, a race condition can occur
when several operations are performed by different processors on a given
area of shared memory, and at least one of these operations is a write.

For instance, if a piece of data located in the shared memory is initialized at
a given value $v_0$ and is accessed concurrently by a process $A$ that reads
this data and a process $B$ that writes the value $v_1$. If $A$ reads it before
$B$ writes, it will read the value $v_0$. If $B$ writes before $A$ reads, $A$
will read $v_1$.

More formally, we can consider read and write operations as \emph{events} in the
distributed system formed by the set of processors and the communication
channels that interconnects them.

Two events $e_1$ and $e_2$ are \emph{ordered} iff there exists an
\emph{happens before} (as defined by~\cite{Lamport78} and denoted~\happensbefore) 
relationship between them such that 
$e_1$\happensbefore $e_2$ or $e_2$\happensbefore $e_1$. Race conditions
are defined in~\cite{Helmbold94a} by the fact that there exists no
causal order between $e_1$ and $e_2$ (further denoted by $e_1\times e_2$).

\ignore{
\emph{Sequentially consistent program}: ``The result of any execution
is the same as if the operations of all the processors were executed
in some sequential order, and the operations of each individual
processor appear in this sequence in the order specified by the
program'' \cite{lamport79}

First requirement: ``Each processor issues memory requests in the
order specified by the program'' -> not sufficient

Second requirement: ``Memory requests from all processors issued to an
individual memory module are serviced from a single FIFO
queue. Issuing a memory request consists of entering the request on
this queue''
}


\Section{Detecting race conditions} 
\label{sec:algo}

In this section, we present an algorithm for detecting race conditions in
parallel applications that follow the distributed shared memory model presented
in section~\ref{sec:model}.

\SubSection{Causal ordering of events}

In section~\ref{sec:model:rc}, we stated that there exists a race condition
between a set of inter-process events when there exists no causal order between
these events. In practice, this definition must be refined: concurrent accesses
that do not modify the data are not problematic. Hence, when an event occurs
between two processes, we need to determine whether it is \emph{causally
 ordered} with the \emph{latest write} on this data.

Lamport clocks~\cite{Lamport78} keep track of the logical time on a process;
vector clocks (introduced by \cite{Mattern88}) allow for the partial causal
ordering of events. A vector clock on a given process contains the logical time
of each other process at the moment when the other process had an influence
on the process (\ie, last time it had a causal influence on this process).

When the causality relationship between a set of events that contains at least a
write event cannot be established, we can conclude that there exists a race
condition between them. More specifically, when we compare the vector clocks
that are associated with these events and the latest write.

\begin{lemma}[Mattern, Theorem 10]
$\forall e,e'\in E : e < e'$ iff $H(e) < H(e')$ and
$e \parallel e'$ iff $C(e) \parallel C(e')$
\end{lemma}

\begin{corollary}
Consider two events denoted $e_1$ and $e_2$ and their respective clocks $H_1$
and $H_2$. If no ordering can be determined between $H_1$ and $H_2$,
there exists a race condition between $e_1$ and $e_2$ ($e_1\times e_2$).
\end{corollary}

In the following algorithms, we detail the \emph{put} and \emph{get}
commands. Algorithm~\ref{algo:algo:put} describes a \emph{put} performed from
$P0$ by the library to write the content of $src$ address into process $P1$'s
memory at address $dst$. Algorithm~\ref{algo:algo:get} describes a \emph{get}
performed by the library to retreive content of $src$ address from process
$P1$'s memory to process $P0$'s memory at address $dst$. Each process associates
two clocks to areas of shared memory: a general-purpose clock $V$ and a
write clock $W$ that keeps track of the latest write operation.

\begin{figure}[htb]
  \centering
  \begin{tikzpicture}[scale=.5]
    \foreach \x in { 0, 1, 2 } {
      \node at ( \x*4, 1 ) {P\x};
      \draw [->] ( \x*4, 0 ) -- ( \x*4, -6 );
    }
    \node at ( -1, 0 ) {a = ?};
    \node at ( 5, 0 ) {a = A};
    \node at ( 9, 0 ) {a = ?};
      
    \draw[thick] ( 0, -1.5 ) -- ( 4, -3.5 );
    \draw[ -latex, thick] ( 4, -3.5 ) -- ( 0, -5.5 );
    \node at ( 1, -3.5 ) {get};
    \node at ( -1, -5.5 ) {a = A};

    \draw[ thick] ( 8, -.5 ) -- ( 4, -2.5 );
    \draw[ -latex, thick] ( 4, -2.5 ) -- ( 8, -4.5 );
    \node at ( 6.5, -2.5 ) {get};
    \node at ( 9, -4.5 ) {a = A};
  \end{tikzpicture}
  \caption{\label{fig:algo:getget}Two concurrent get operations}
\end{figure}
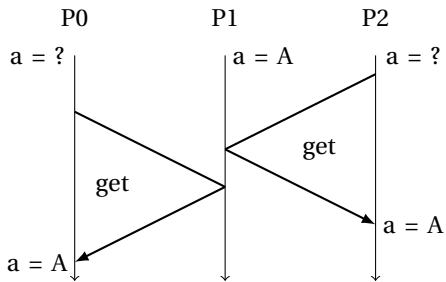

Figure~\ref{fig:algo:getget} shows an example of two concurrent remote read
operations (\ie, \emph{get} operations) on a variable $a$. This variable is
initialized at a given value $A$ before the remote accesses. Since none of the
concurrent operations modifies its value, this is not a race condition. As stated
in section~\ref{sec:model:rc}, there exists a race condition between concurrent
data accesses iff at least one access modifies the value of the data. As a
consequence, concurrent read-only accesses must not be considered as race
conditions. 

\begin{algorithm}[ht]
  \Begin{
    lock$( P0, src )$\;
    lock$( P1, dst )$\;
    $V$ = update\_local\_clock$( P0, src )$\;
    $W'$ = get\_clock\_W$( P1, src )$\;
    \If{ $\neg$ compare\_clocks$( V, V' )$ \\
      \ \ $\wedge \neg$ compare\_clocks$( V', V )$ }{
      signal\_race\_condition() \;
    }
    put$( P0, src, P1, dst )$\;
    update\_clock\_W$( P1, dst )$\;
    update\_clock$( P1, dst )$\;
    unlock$( P1, dst )$\;
    unlock$( P0, src )$\;
  }
  \caption{\emph{Put} operation from $P0$ to $P1$}
  \label{algo:algo:put}
\end{algorithm}

\begin{algorithm}[ht]
  \Begin{
    lock$( P0, dst )$\;
    lock$( P1, src )$\;
    $V$ = update\_local\_clock$( P0, dst )$\;
    $W = V$
    $V'$ = get\_clock$( P1, src )$\;
    \If{ $\neg$ compare\_clocks$( W, V' )$ \\
      \ \ $\wedge \neg$ compare\_clocks$( V', V )$ }{
      signal\_race\_condition() \;
    }
    get$( P1, src, P0, dst )$\;
    update\_clock$( P1, src )$\;
    update\_clock$( P0, dst )$\;
    unlock$( P1, dst )$\;
    unlock$( P0, dst )$\;
  }
  \caption{\emph{Get} operation from $P0$ to $P1$}
  \label{algo:algo:get}
\end{algorithm}

The $lock$ primitive takes care of mutual exclusion if the addressed value is in
public space or 
not. If the address is in private space, there is no need of a real lock (except
in multithreading). The $compare\_clocks(P0,a,P1,b)$ primitive first read the
vector clock $V_1(b)$ from $P1$'s memory and then compare it with $V_0(a)$. The
comparison is done as described in algorithm~\ref{algo:algo:compare}. 

\begin{algorithm}[H]
  \Begin{
    return $\forall n \in \{0, \dots, N-1\}$ : $V_{Pi} < V_{Pj} \Leftrightarrow $ \\
    \hfill $V_{Pi}[n] < V_{Pj}[n]$ ) \;
  }
  \caption{compare\_clocks algorithm}
  \label{algo:algo:compare}
\end{algorithm}

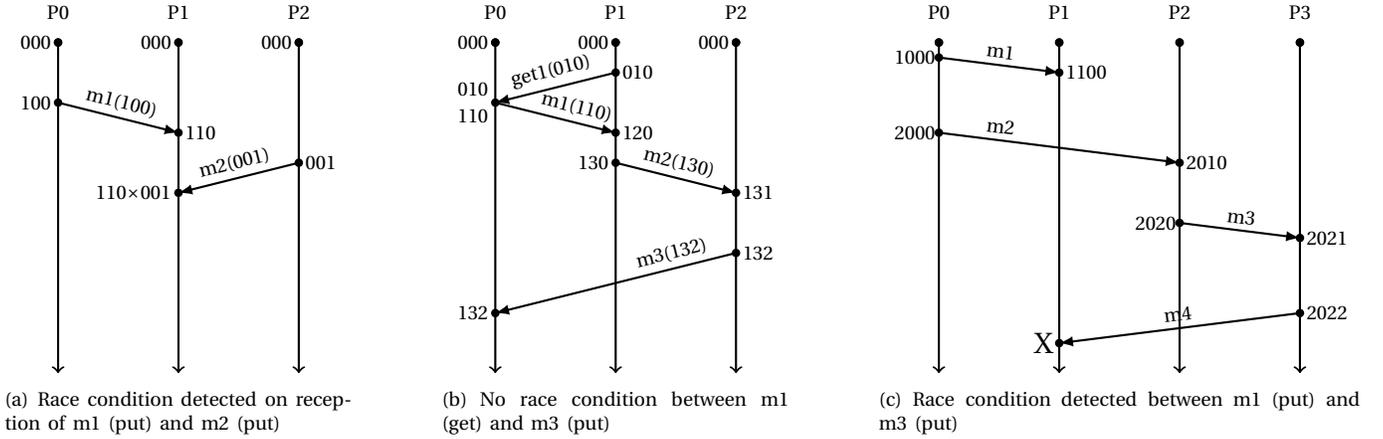
\begin{figure*}[htb]
  \subfigure[Race condition detected on reception of m1 (put) and m2 (put)]{
    \begin{tikzpicture}[scale=.4]
      \footnotesize
      \chronogramme{2}
	\node at (PP0)[left] {000};
	\node at (PP1)[left] {000};
	\node at (PP2)[left] {000};
	\noeud{(0,-2)}{departm1}
	\noeud{(4,-3)}{arrivm1}
      \path[frame] (departm1)  -- (arrivm1) node[midway,above,sloped]{m1(100)};
      \node at (departm1)[left] {100};
      \node at (arrivm1)[right] {110};
	\noeud{(8,-4)}{departm2}
	\noeud{(4,-5)}{arrivm2}
      \path[frame] (departm2) -- (arrivm2) node[midway,above,sloped]{m2(001)};
      \node at (departm2)[right] {001};
      \node at (arrivm2)[left] {110$\times$001};
    \end{tikzpicture}
    \label{fig:algo:vector:exemple1}
  }\hfill
  \subfigure[No race condition between m1 (get) and m3 (put)]{
    \begin{tikzpicture}[scale=.4]
      \footnotesize
      \chronogramme{2}
	\node at (PP0)[left] {000};
	\node at (PP1)[left] {000};
	\node at (PP2)[left] {000};
	\noeud{(4,-1)}{departget1}
	\noeud{(0,-2)}{arrivget1}
      \path[frame] (departget1) -- (arrivget1) node[midway,above,sloped]{get1(010)};
      \node at (departget1)[right] {010};
      \node at (arrivget1)[left, anchor=south east] {010};
      \node at (arrivget1)[left, anchor=north east] {110};
	\noeud{(4,-3)}{arrivm1}
      \path[frame] (arrivget1) -- (arrivm1) node[pos=0.65,above,sloped]{m1(110)};
      \node at (arrivm1)[right] {120};

	\noeud{(4,-4)}{departm2}
	\noeud{(8,-5)}{arrivm2}
      \node at (departm2)[left] {130};
      \path[frame] (departm2) -- (arrivm2) node[midway,above,sloped]{m2(130)};
	\node at (arrivm2)[right] {131};

	\noeud{(8,-7)}{departm3}
	\noeud{(0,-9)}{arrivm3}
      \path[frame] (departm3) -- (arrivm3) node[near start,above,sloped]{m3(132)};
      \node at (departm3)[right] {132};
      \node at (arrivm3)[left] {132};
    \end{tikzpicture}
    \label{fig:algo:vector:exemple2}
  }\hfill
  \subfigure[Race condition detected between m1 (put) and m3 (put)]{
   \begin{tikzpicture}[scale=.4]
      \footnotesize
      \chronogramme{3}
        \noeud{(0,-0.5)}{departm1}
        \noeud{(4,-1)}{arrivm1}
      \path[frame] (departm1) -- (arrivm1) node[midway,above,sloped] {m1};
      \noeudMatrice{departm1}{left}{1000}
      \noeudMatrice{arrivm1}{right}{1100}

      \noeud{(0,-3)}{departm2}
      \noeud{(8,-4)}{arrivm2}
      \path[frame] (departm2) -- (arrivm2) node[near start,above,sloped] {m2};
      \noeudMatrice{departm2}{left}{2000}
      \noeudMatrice{arrivm2}{right}{2010}

      \noeud{(8,-6)}{departm3}
      \noeud{(12,-6.5)}{arrivm3}
      \path[frame] (departm3) -- (arrivm3) node[midway,above,sloped] {m3};

      \noeudMatrice{departm3}{left}{2020}
      \noeudMatrice{arrivm3}{right}{2021}
      \noeud{(12,-9)}{departm4}
      \noeud{(4,-10)}{arrivm4}
      \path[frame] (departm4) -- (arrivm4) node[midway,above,sloped] {m4};

      \noeudMatrice{departm4}{right}{2022}
        \node at (arrivm4)[left]{\large X};
    \end{tikzpicture}
    \label{fig:algo:vector:exemple3}
  }
  \hfill
  \caption{\label{fig:algo:vector}Detecting race conditions with vector clocks}
\end{figure*}

In figure~\ref{fig:algo:vector}, we present three use-cases of our algorithm:
two situations of race conditions and one when the messages are causally ordered.

\SubSection{Clock update}

The clock matrix $V_{Pi}$ is maintained by each process $Pi$. This matrix is a
\emph{local} view of the global time. It is initially set to zero. Before $Pi$
performs an event, it increments its local logical clock $V_{Pi}[i, i]$
($update\_local\_clock$). Clocks are updated by any event as follows
(algorithm~\ref{algo:algo:maxclock}, see~\cite{Raynal96}).

\begin{algorithm}[H]
  \Begin{
    $\forall l, V'[l] = \max( V_{Pi}[l], V_{Pj}[l] )$\;
    return V' \;
  }
  \caption{max\_clock algorithm}
  \label{algo:algo:maxclock}
\end{algorithm}
 
The remote clock update is performed as follows:

\begin{algorithm}[H]
  \Begin{
    $V_{Pj}$ = get\_clock$( P_j, dst )$\;
    $V'$ = max\_clock$( V_{Pi}, V_{Pj} )$\;
    put\_clock$( P_j, dst, V' )$\;
  }
  \caption{update\_clock algorithm}
  \label{algo:algo:updateclock}
\end{algorithm}

The update\_clock\_W algorithm is similar to the update\_clock algorithm, except
that it updates the value of the ``write clock'' $W$.

Since the shared memory area is locked, there cannot exist a race condition
between the remote memory accesses induced by the race condition detection
mechanism.

\SubSection{Discussion on the size of clocks}
\label{algo:discutaille}

If $n$ denotes the number of processes in the system, it has been shown that the
size of the vector clocks must be at least $n$~\cite{CB91}. As a consequence,
the size of the clocks cannot be reduced.

\SubSection{Discussion on error signalisation}

A race condition may not be fatal: some algorithms contain race conditions on
purpose. For example, parallel master-worker computation patterns induce a race
condition between workers when the results are sent to the master. Therefore,
race conditions must be \emph{signaled} to the user (\eg, by a message on the
standard output of the program), but they must not abort the execution of the
program.

In the algorithm presented here, we refine the error detection by using two
distinct clocks, a general-purpose one and a ``write clock''. The drawback of
this approach is that it doubles the necessary amount of memory. On the other
end, it offers more precision and eliminates numerous cases of false positives
(\eg, concurrent read-only accesses).


\Section{Conclusion and perspective}

In this paper, we presented a model for distributed shared memory. This model
considers interactions between processes and causal dependencies, while taking
into account specific features from hardware used to implement such systems. 

In this model, we propose an algorithm for detecting race conditions caused by
the absence of ordering between events in the distributed system. This algorithm
can be implemented in the communication library of the run-time support system
that executes the program on a distributed system.

\SubSection{Discussion}

As stated in section~\ref{algo:discutaille}, the size of the matrices cannot be
smaller than $n$, if $n$ denotes the number of processes in the
system. Moreover, a clock must be used for each shared piece of data. As a
consequence, our algorithm has an overhead on data storage space (clocks
associated with shared data) and with communication performance. However, race
condition detection is typically a \emph{debugging} technique. It does not need
to be enabled on a parallel application that is actually running at full
performance and large-scale systems. Parallel programmes are typically debugged
on small data sets and a few processes (typically, about 10 processes). 

\SubSection{Future works}

The model presented in this paper leads to new interpretations of distributed
algorithms. New operations can also be imagined, such as non-collective, global
operations: for example, a process can perform a reduction (\ie, a global
operation on some data held by all the other processes) without any
participation for the other processes, by fetching the data remotely.

Our race condition detection algorithm can be implemented at two levels: in the
communication library of a parallel language, for automatic detection of
conflictual accesses, or in the pre-compiler, as wrappers around remote
data accesses.


\bibliographystyle{IEEEtran}
\IEEEtriggeratref{19}
\bibliography{biblio}

\end{document}